# Uncovering the ultimate performance of single-walled carbon nanotube films as transparent conductors

K. Mustonen[1], P. Laiho[1], A. Kaskela[1], T. Susi[2], A.G. Nasibulin[1,3], E.I. Kauppinen[1,a)]

[1] *Department of Applied Physics, Aalto University School of Science, P.O. Box 15100, FI-00076 Aalto, Finland*

[2] *University of Vienna, Faculty of Physics, Boltzmanngasse 5, A-1090 Vienna, Austria*

[3] *Skolkovo Institute of Science and Technology, 100 Novaya st., Skolkovo, Odintsovsky district, Moscow Region, 143025, Russia*

---

[a)] Electronic mail: esko.kauppinen@aalto.fi

The ultimate performance – ratio of electrical conductivity to optical absorbance – of single-walled carbon nanotube (SWCNTs) transparent conductive films (TCFs) is an issue of considerable application relevance. Here, we present direct experimental evidence that SWCNT bundling is detrimental for their performance. We combine floating catalyst synthesis of non-bundled, high-quality SWCNTs with an *aggregation chamber*, in which bundles with mean diameters ranging from 1.38 to 2.90 nm are formed from identical 3 μm long SWCNTs. The as-deposited TCFs from 1.38 nm bundles showed sheet resistances of 310 Ω/□ at 90% transparency, while those from larger bundles of 1.80 and 2.90 nm only reached values of 475 and 670 Ω/□, respectively. Based on these observations, we elucidate how networks formed by smaller bundles perform better due to their greater interconnectivity at a given optical density. Finally, we present a semi-empirical model for TCF performance as a function of SWCNT mean length and bundle diameter. This gives an estimate for the ultimate performance of non-doped, random network mixed-metallicity SWCNT TCFs at ~80 Ω/□ and 90% transparency.

Single-walled carbon nanotubes (SWCNTs) are an appealing material for replacing widely used metal oxides as transparent conducting films (TCFs) due to their relatively high conductivity, excellent flexibility and low refractive index.[1] The superb conductivity of individual nanotubes originates from the exceptionally high charge carrier mobility in the SWCNT quantum channel, allowing them to carry current densities one thousand times higher than copper.[2] Nevertheless, the sheet conductivity of even dense networks[3] of high-quality SWCNTs falls far short of expectations based on the electronic properties of individual tubes.[4] This disparity has led to a consensus that the conductivity of SWCNT networks is not limited by the resistance of the nanotubes, but rather by the barriers, or junction resistances, between them.[5-8] The other defining factor in TCF performance, the absorption of visible light, follows the Beer-Lambert law and is directly proportional to the number of carbon atoms in the tubes' graphitic lattices,[9] along a minor contribution from optical transitions of metallic tubes that fall in the visible range.

Hence, the issue of TCF performance can be approached as an optimization problem for the ratio $K$ between the sheet conductance ($\sigma_{DC}$), and light absorbance ($A$), usually measured for 550 nm light:

$$K = \sigma_{DC}/A = 1/(R_s \, log_{10}[T]), \qquad (1)$$

where the second equality is an alternative expression using transmittance ($T$) and sheet resistance ($R_s$). The goal therefore is to achieve the highest possible $\sigma_{DC}$ with as little absorption, and thus as few carbon atoms, as possible. We have earlier postulated,[10] together with Lyons *et al.* and Hecht *et al.*,[3,11] that this can be achieved by minimizing the nanotube average bundle diameter $\langle d_b \rangle$, maximizing the number of parallel conduction paths. So far these proposals have remained speculative, since available samples have differed also in SWCNT lengths and other confounding factors, which could not be completely accounted for.[3,10-12]

Here, we directly test the bundle diameter effect by combining our recently developed floating catalyst chemical vapour decomposition (FC-CVD) synthesis route of intrinsically non-bundled SWCNTs[13] with an *aggregation chamber*, in which the nanotubes are allowed to controllably bundle. Inside the chamber, tube-tube collisions and gas phase aggregation (bundling) are caused by Brownian diffusion.[14] Hence, the time dependence of $\langle d_b \rangle$ in the gas-phase is governed by diffusion theory and, assuming monodisperse aggregation, the aerosol number concentration evolves in time as[14]

$$N(t) = {N_0}/{(1 + N_0 K_\phi(D_M, T, \eta)t)}, \qquad (2)$$

where $K_\phi$ is the aggregation coefficient depending on particle mobility diameter ($D_M$), surrounding gas temperature ($T$) and viscosity ($\eta$), and $N_0$ is the number concentration at $t_0$.

The average number of nanotubes per bundle $\langle n(t) \rangle$, increasing by aggregation due to the collisions described by Equation 2, can be expressed as

$$\langle n(t) \rangle = \langle n_0 \rangle + \left({N_0}/{N(t)}\right) - 1 = \langle n_0 \rangle + N_0 K_\phi t, \qquad (3)$$

where $\langle n_0 \rangle$ is the number of nanotubes at $t_0$, and the term -1 is necessary for consistency with Equation 2. Assuming an average tube diameter $\langle d_{CNT}\rangle$, geometrically $\langle d_b\rangle \approx \langle d_{CNT}\rangle\sqrt{\langle n\rangle}$.[11] When colliding, SWCNTs pack closely due to the van der Waals interaction,[15] and thus the evolution of $\langle d_b\rangle$ in time can be approximated as

$$\langle d_b(t)\rangle \approx \langle d_{CNT}\rangle\sqrt{\langle n(t)\rangle} = \langle d_{CNT}\rangle\left(\sqrt{\langle n_0\rangle} + K_\phi N_0 t\right). \qquad (4)$$

To achieve a predictable increase in bundle diameter, the aggregation chamber provides a residence time of 360 s under a laminar flow of 500 cm³ min⁻¹ (Reynolds number ~0.06, Figure 1). The chamber has a circular cross-section with a diameter of 6 cm, total length of 120 cm, and is connected to a ¼" stainless steel line with two conical end-pieces to maintain turbulence-free gas flow. Most of the nanotube aggregation takes place in the chamber, but at higher number concentrations, some bundles are also formed inside the synthesis reactor.[13]

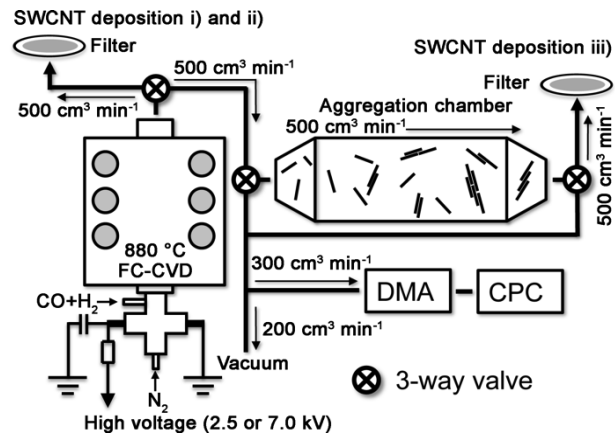

Fig 1. Experimental system for the synthesis and aggregation of SWCNTs, and TCF deposition. The catalyst particles were formed in a spark discharge generator[13] and fed to the FC-CVD reactor at different number concentrations (*N*) to form SWCNTs. The nanotubes were either directly collected at reactor outlet (spark voltage of 2.5 kV or 7 kV), or, additionally, at the 7 kV voltage passed through the aggregation chamber to promote bundle

growth prior to deposition. A combination of a DMA and a CPC was used to acquire number size distributions (NSDs).

The SWCNTs were grown in the FC-CVD reactor (Figure 1) heated to 880 °C in a flow of carbon monoxide (CO), hydrogen ($H_2$) and nitrogen ($N_2$) at mole fractions of 0.7, 0.1 and 0.2, respectively.[13] The total volumetric flow rate through the reactor was 500 $cm^3 min^{-1}$. Individual tubes were synthesized at low catalyst concentration, i.e. at a spark recharge voltage of 2.5 kV, effectively inhibiting bundling during synthesis. A higher number concentration was achieved by setting the voltage to 7 kV, leading to slight bundle formation also during synthesis. Samples were thus deposited from three conditions: **i)** Minimizing $\langle d_b \rangle$ during synthesis using a recharge voltage of 2.5 kV, followed by deposition directly at the reactor outlet; **ii)** Increasing $\langle d_b \rangle$, taking place during synthesis using 7 kV, followed by direct deposition; **iii)** Maximizing $\langle d_b \rangle$ by synthesis using 7 kV, followed by deposition after an additional residence time of 360 seconds in the aggregation chamber. The number size distributions (NSDs) from different conditions were measured using a combination of a differential mobility analyser (DMA, Vienna-type) and a condensation particle counter (CPC model 5414, Grimm Aerosol Technik GmbH, Germany). Figure 2a shows the NSDs, indicating that the number concentration increases almost 5-fold (from $2.35 \times 10^5$ $cm^{-3}$ to $1.14 \times 10^6$ $cm^{-3}$) by increasing the operating voltage from 2.5 kV to 7.0 kV. After an additional 360 s of bundling in the aggregation chamber, the collisions again reduce the number concentration down to $3.24 \times 10^5$ $cm^{-3}$, clearly indicating the effect of aggregation. The electrical mobility geometric mean diameter, $\langle D_{ME} \rangle$, related to the physical mobility of floating bodies under the influence of a static electric field, concurrently increases from 33 to 41 nm from condition **i)** to **iii)**.

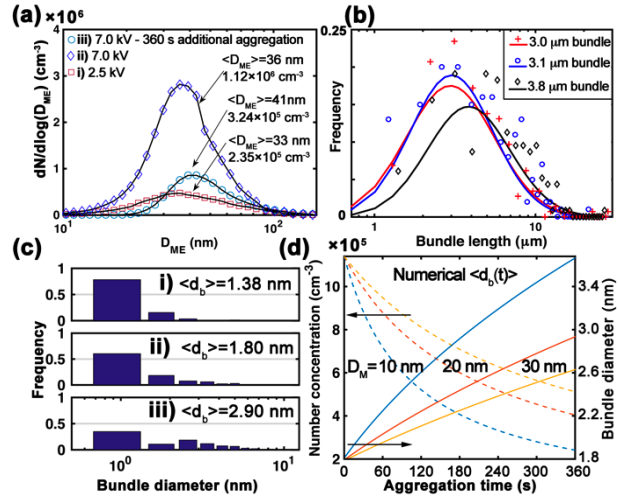

Fig 2. **(a)** The SWCNT number size distributions (NSDs) as measured with a combination of DMA and CPC from conditions **i-iii) (b)** Bundles deposited from condition **i)** exhibit a mean length $\langle L_{bundle} \rangle$ of 3.0 μm, whereas conditions **ii)** and **iii)** result in slightly longer bundles of 3.1 and 3.8 μm, respectively. **(c)** Correspondingly, bundles from condition **i)** exhibit a mean bundle diameter $\langle d_b \rangle$ of 1.38 nm, whereas conditions **ii)** and **iii)** yield 1.80 and 2.9 nm, respectively. **(d)** Numerically solved Equations 2 and 4 provide $N(t)$ and $\langle d_b(t) \rangle$ using mean tube diameter $\langle d_{CNT} \rangle$ of 1.1 nm[13] and mobility diameters ($D_M$) of 10, 20 and 30 nm at ambient conditions.

This NSD analysis is corroborated by the $\langle d_b \rangle$ statistics (Figure 2c), with an increasing trend with higher *N* and bundling time. The statistics shown were gathered using an atomic force microscope (AFM, Veeco Dimension 5000, Switzerland; operated in tapping mode), showing a significant increase in the mean bundle diameter from 1.38 to 2.90 nm from condition **i)** to **iii)**. Considering that the mean tube diameter $\langle d_{CNT} \rangle$ is 1.1 nm as shown previously,[13] the largest bundles contain more than four times more carbon per unit length than the smallest ones. For AFM observations, the SWCNTs were deposited on mica substrates using a custom thermophoretic precipitator[16]; representative micrographs can be found in Ref 17. To draw conclusions from the SWCNT mobility diameter, $D_M$, Equations 2 and 4 were numerically solved using the initial values $N=1.12\times10^6$ cm$^{-3}$ from the DMA+CPC measurement and $\langle d_b \rangle$ = 1.80 nm from the AFM measurements. The calculation was separately run for $D_M$ = 10, 20 and 30 nm, of which 20 nm provided the closest match to observations with $\langle d_b \rangle$ = 2.91 nm and $N$ = 4.04×10$^5$ cm$^{-3}$, as depicted in Figure 2d. The predicted bundle diameter closely matches our observations, but the final $N$ is over-estimated by ~20%. This discrepancy likely arises from the omission of particle diffusion to the chamber walls from the calculation.

Importantly, TCF performance can only be comprehensively analysed with respect to $\langle d_b \rangle$ when the SWCNT lengths $\langle L_{CNT} \rangle$ are invariant from sample to sample. The bundle length statistics shown in Figure 2b were gathered using a scanning electron microscope (SEM) (Zeiss Sigma VP, Carl Zeiss AG, Germany) to image SWCNTs deposited on Si/SiO$_2$ chips using the thermophoretic precipitator. Between conditions **i)** and **ii)**, $\langle L_{bundle} \rangle$ remains relatively unchanged, (3.0 μm and 3.1 μm, respectively) whereas in condition **iii)** the apparent bundle length increases to 3.8 μm. Since the temperature, gas composition, and residence time in the synthesis conditions are identical, the observed slight increase in $\langle L_{bundle} \rangle$ between conditions **i)** and **iii)** must arise solely from the formation of bundles.

For optical and electrical characterization, SWCNT films were filtered from the gas flow onto membrane filters (pore size 0.47 μm, Millipore, France) and press-transferred[12] onto clean quartz substrates. The TCF thickness was varied by adjusting the deposition time and the transmittance at 550 nm determined using a UV/Vis-NIR spectrophotometer (PerkinElmer Inc. Lambda 950, USA), in which the substrate contribution was subtracted via a clean reference sample in the secondary beam path. The sheet resistances ($R_S$) were determined using a 4-point probe (Jandel Engineering Ltd., UK) at a 60 g needle loading.

The sheet resistances vs. transmittances of SWCNT films deposited from conditions **i)-iii)** are shown in Figure 3, together with trendlines fitted according to Equation 1. Remarkably, these show that the films' conductivity at a certain optical density has an inverse relation to $\langle d_b \rangle$: as-deposited TCFs fabricated from 1.38 nm bundles achieve a much higher as-deposited performance, 310 Ω/□ at $T_{550nm}$=90% ($K$=32.80 kΩ$^{-1}$), than TCFs deposited from larger 1.80 and 2.90 nm bundles, which reach values of 475 and 670 Ω/□ ($K$=21.90 kΩ$^{-1}$ and 15.50 kΩ$^{-1}$), respectively. If the largest diameter sample would actually contain longer tubes, this should decrease its resistance compared to the other two conditions, which is not seen. As additional supporting evidence of bundling, a 20 nm redshift is observed for the $E_{ii}$ optical transitions ($E_{11}$ in Figure 3 inset, the rest in Ref 17) between conditions **i)** and **iii)**. According to Wang *et al.,* such a redshift can be attributed to increased dielectric screening in bundled tubes, slightly decreasing the exciton lifetimes through intra-bundle coupling.[18]

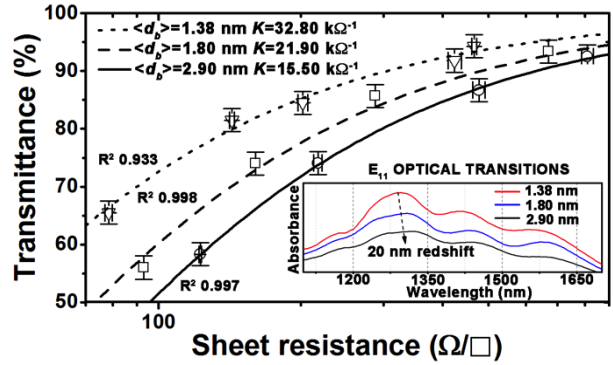

Fig 3. Transmittance vs. sheet resistance for TCFs fabricated from SWCNTs with different bundle diameters $\langle d_b \rangle$; trendlines fitted according to Equation 1. The smallest 1.38 nm bundles give the lowest $R_s$ of 310 Ω/□ at $T_{550\,nm}$=90%. By comparison, larger bundles yield only 475 Ω/□ and 670 Ω/□ at $T_{550nm}$=90%. The lower right inset shows the observed redshift of the $E_{11}$ optical transitions.

Since all TCFs in this work were fabricated from identical SWCNTs, the performance variation must emerge solely from the organization of the nanotubes in the network – larger diameter bundles simply absorb more than individual tubes at the same conductivity. This implies that larger bundles do not conduct much better to compensate for their increased absorption. This could be attributed to least a few different factors: Delaney *et al.* and Oyuang *et al.* reported the formation of pseudo band gaps in bundles of metallic tubes, which would increase barrier heights between bundles and directly impact the conductivity of the junctions.[19] This, however, seems unlikely here, since Znidarsic *et al.* reported the opposite effect using conductive tip AFM (C-AFM) of bundles deposited with a method similar to ours.[7] A more likely explanation is provided by the anomalous conduction effect reported by Radosavljević *et al.* for mixed-metallicity bundles and Han *et al.* for bundles of semiconducting tubes.[20] Their observations suggest that charge carrier transport mainly occurs on the outermost layer of tubes, leaving the bundle core as effectively "dead mass", contributing only to optical absorption. In agreement with this argument, the mentioned C-AFM study found only modest variation, between 3 and 16 kΩ μm$^{-1}$, in the length-wise resistances of bundles of different sizes. Previously, we proposed that when possible variations in junction resistances with $\langle d_b \rangle$ are not accounted, larger bundles should be detrimental for the TCF quality through increasing absorption.[10] However, we were previously unable to simultaneously maintain SWCNT lengths, adding a source of uncertainty to the analysis.

To refine this idea, we define an '*ideal quality factor*' $K_0$ of SWCNT TCFs, which is the quality factor $K$ of a network fabricated from 100% individual tubes ($\langle n \rangle = 1$). Due to absorption being related to the amount of carbon, we find $K \propto \langle d_b \rangle^{-2} \approx \langle d_{CNT} \rangle^{-2} \langle n \rangle^{-1}$. As some bundling will always occur at least during deposition, $K_0 > K$, making $K/K_0(n)$ a monotonically decreasing function, with a maximum of $K/K_0(1) = 1$. For the 3 μm tubes presented in this study, $K_0$ can be approximated by assuming $K$ and $\langle n \rangle$ are roughly linearly proportional near $\langle n \rangle = 1$, giving $K_0$= 34.2±3.5 kΩ$^{-1}$ (details in Ref 17).

Thus, while by definition $K/K_0$ decreases with increasing $\langle n \rangle$, their exact relationship needs more clarification. As we mentioned earlier, junction resistances $R_{jct}$ seem to be lower for bundles than for individual tubes.[7] Based on the seminal work of Fuhrer *et al.*,[6] we know that junctions between metallic tubes may exhibit resistances orders of magnitudes lower than junctions between mixed or semiconducting tubes. In a dense random network of SWCNTs, we can thus assume for the purposes of the following formulation that current conduction is dominated by metallic pathways. The probability of forming a junction between two metallic tubes or bundles is a function of the metallic-to-semiconducting ratio (1:2 in typical CVD processes) and the average number of tubes per bundle $\langle n \rangle$.[21] This can be written as $P_M(n) = 1 - 2/3^{\langle n \rangle}$, which is the probability of one tube in the bundle being metallic (derivation presented in Ref 17). Taking into account that $R_s$ is dominated by junction resistances, and that $K \propto R_s^{-1}$ (Equation 1), an alternate definition for the quality factor $K$ using $\langle n \rangle$ and $K_0$ can be expressed as

$$K = K_0 \beta \left(1/\langle n \rangle\right)\left(1 - [2/3]^{\langle n \rangle}\right), \tag{5}$$

where $\beta$ is a dimensionless fitting parameter. Figure 4b plots the prediction provided by $K/K_0 = (\beta/\langle n \rangle)(1 - [2/3]^{\langle n \rangle})$, set to intersect the point $K/K_0(1) = 1$. Equation 5 fits very well to the data with the parameter $\beta = 2.67$.

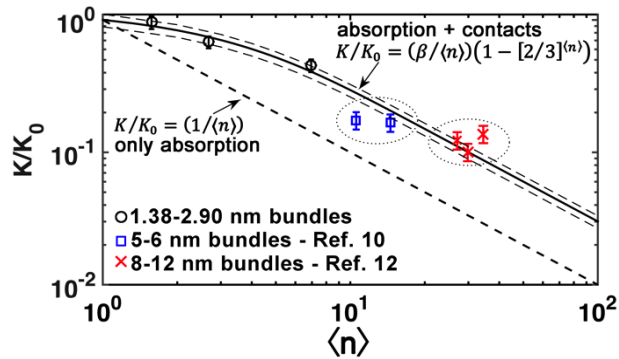

Fig 4. The scaling of $K/K_0$ as a function of $\langle n \rangle$ according to Equation 5 shows that larger bundles provide lower performance compared to small bundles or individual tubes. The black circles are data from this work, and the blue and red are estimates from Refs 10 and 12.

Finally, we compare the prediction of Equation 5 with our earlier published results. We chose to use our own data since both the bundle lengths and diameters were known and, due to the same fabrication method, we can assume the tubes are mostly unaffected by contaminants such as surfactants.[10,12] However, since the SWCNT lengths and $\langle n \rangle$ were not directly known, those needed to be approximated on available data. $\langle n \rangle$ was estimated using the relation $\langle d_b \rangle \approx \langle d_{CNT} \rangle \sqrt{\langle n \rangle}$, while the SWCNT lengths were estimated using the published bundle diameters and lengths, with the approximations: $\langle d_b \rangle$ = 3–5 nm → $\langle L_{CNT} \rangle = 3/4 \langle L_{bundle} \rangle$; $\langle d_b \rangle$ = 5–9 nm → $\langle L_{CNT} \rangle = 3/5 \langle L_{bundle} \rangle$; and $\langle d_b \rangle$ = 9–12 nm → $\langle L_{CNT} \rangle = 1/2 \langle L_{bundle} \rangle$. (The uncertainty of this approximation is estimated to be about 20%.) Based on previous studies, SWCNT mean lengths and sheet conductivity relate linearly, $\sigma_{DC} \propto K \propto \langle L_{CNT} \rangle$.[3,22] The *ideal quality factors* for SWNCTs of any length can thus be calculated as $K_0' = K_0(\langle L_{CNT}' \rangle / \langle L_{CNT\ 0} \rangle)$, where $\langle L_{CNT\ 0} \rangle$ and $K_0$ are 3 μm and 34.2 kΩ$^{-1}$, respectively. These additional estimated data points are plotted in Figure 4, where we see that the model fits the additional data well.

We can thus provide a semi-empirical formula that can be used to predict TCF performance: by combining the relation $K \propto \langle L_{CNT} \rangle$ with Equation 5, the quality factor can be expressed in terms of $\langle n \rangle$ and $L_{\langle L_{CNT} \rangle}$ as

$$K(n, L_{CNT}) = \Pi \left( \frac{(1-[2/3]^{\langle n \rangle})\langle L_{CNT} \rangle}{\langle n \rangle} \right), \qquad (6)$$

where $\Pi = \beta K_0 / \langle L_{CNT\ 0} \rangle$ =30.4±5.8 μm$^{-1}$ kΩ$^{-1}$. However, increasing $\langle L_{CNT} \rangle$, which at constant optical density corresponds to decreasing the junction density, can only improve the performance until the lengthwise resistances between junctions reach the same magnitude as the junction resistances themselves.[23] According to Purewal *et al.* and Znidarsic *et al.*, who provide direct measurements of these quantities, that point is reached somewhere between 10 and 20 μm.[7,8] Importantly, this will also set the ultimate conduction limit for pristine, mixed-metallicity percolating SWCNT networks; should the scaling law apply until, say, a conservative estimate of $\langle L_{CNT} \rangle$=10 μm, the highest achievable quality factor would be roughly 114 kΩ$^{-1}$, corresponding to 80±15 Ω/□ at 90% transparency (assuming our model holds, an all-metallic network could reach ~25 Ω/□). Treatment with strong chemical dopants such as nitric acid ($HNO_3$) may be used to push this limit further, though how much improvement can be achieved likely depends on the network morphology, chiral distribution and the selected dopant.

To summarize, we have by experimental design demonstrated the detrimental influence single-walled carbon nanotube bundling has on the performance of transparent conducting films. We explain the decrease in sheet conductivity at a certain optical density through geometric factors: current transport mainly takes place on the surfaces of bundles, while the nanotubes inside the bundles continue to contribute to light absorption. In addition, based on our experimental results and analysis, we have formulated a semi-empirical model that predicts the optimal TCF conductivity at a certain optical density (80±15 Ω/□ at 90% transparency) for arbitrarily large bundles, and nanotube lengths up to 10-20 μm. No SWCNT TCF reported in the literature has exceeded this limit, which we suggest is the ultimate goal for process optimization.

**Acknowledgements**

The research leading to these results has received funding from the European Union Seventh Framework Programme (*FP7/2007-2013*) under *grant agreement* n° 604472 (IRENA project) and n° 314068 (TREASORES project), by the Aalto Energy Efficiency (AEF) program through the MOPPI project, and from TEKES projects CARLA and USG and Academy of Finland (HISCON and 276160). A.G.N. was partially supported by the Ministry of Education and Science of the Russian Federation (Project DOI: RFMEFI58114X0006), and T.S. by the Austrian Science Fund (FWF) through grant M 1497-N19, by the Finnish Cultural Foundation, and by the Walter Ahlström Foundation. This work made use of the Aalto University Nanomicroscopy Center (Aalto-NMC) premises. The personnel of the National Nanomicroscopy Center of Aalto University are gratefully acknowledged for useful discussions and assistance.

**References**

1. Z. Wu, Z. Chen, X. Du, J. M. Logan, J. Sippel, M. Nikolou, K. Kamaras, J. R. Reynolds, D. B. Tanner, A. F. Hebard, and A. G. Rinzler, Science **305** (5688), 1273 (2004).

2. H. Seunghun and M. Sung, Nat Nano **2** (4), 207 (2007).

3. P. E. Lyons, D. Sukanta, F. Blighe, V. Nicolosi, L. F. C. Pereira, M. S. Ferreira, and J. N. Coleman, Journal of Applied Physics **104** (4), 044302 (2008).

4. T. W. Ebbesen, H. J. Lezec, H. Hiura, J. W. Bennett, H. F. Ghaemi, and T. Thio, Nature **382** (6586), 54 (1996);   H. Dai, E. W. Wong, and C. M. Lieber, Science **272** (5261), 523 (1996).

5. P. N. Nirmalraj, P. E. Lyons, S. De, J. N. Coleman, and J. J. Boland, Nano Letters **9** (11), 3890 (2009).

6. M. S. Fuhrer, J. Nygård, L. Shih, M. Forero, Y. Yoon, M. S. C. Mazzoni, H. J. Choi, J. Ihm, S. G. Louie, A. Zettl, and P. L. McEuen, Science **288** (5465), 494 (2000).

7. A. Znidarsic, A. Kaskela, P. Laiho, M. Gaberscek, Y. Ohno, A. G. Nasibulin, E. I. Kauppinen, and A. Hassanien, The Journal of Physical Chemistry C **117** (25), 13324 (2013).

8. M. S. Purewal, B. H. Hong, A. Ravi, B. Chandra, J. Hone, and P. Kim, Physical Review Letters **98** (18), 186808 (2007).

9. J. Blancon, M. Paillet, H. N. Tran, X. T. Than, S. A. Guebrou, A. Ayari, A. S. Miguel, N. Phan, A. Zahab, J. Sauvajol, N. D. Fatti, and F. Vallée, Nat Commun **4** (2013).

10. K. Mustonen, T. Susi, A. Kaskela, P. Laiho, Y. Tian, A. G. Nasibulin, and E. I. Kauppinen, Beilstein Journal of Nanotechnology **3**, 692 (2012).

11. D. S. Hecht, L. Hu, and G. Gruner, Applied Physics Letters **89** (13), 133112 (2006).

12. A. Kaskela, A. G. Nasibulin, M. Y. Timmermans, B. Aitchison, A. Papadimitratos, Y. Tian, Z. Zhu, H. Jiang, D. P. Brown, A. Zakhidov, and E. I. Kauppinen, Nano Letters **10** (11), 4349 (2010).

13. K. Mustonen, P. Laiho, A. Kaskela, Z. Zhu, O. Reynaud, N. Houbenov, Y. Tian, T. Susi, H. Jiang, A. G. Nasibulin, and E. I. Kauppinen, Applied Physics Letters **107** (1), 013106 (2015).

14. W. C. Hinds, *Aerosol Technology: Properties, Behavior, And Measurement of Airborne Particles*. (Wiley-Interscience, 1999).

15. S. Cranford, H. Yao, C. Ortiz, and M. J. Buehler, Journal of the Mechanics and

Physics of Solids **58** (3), 409 (2010).

[16] D. Gonzalez, A. G. Nasibulin, A. M. Baklanov, S. D. Shandakov, D. P. Brown, P. Queipo, and E. I. Kauppinen, Aerosol Science and Technology **39** (11), 1064 (2005).

[17] See supplemental material at [URL will be inserted by AIP] for AFM and SEM micrographs and optical absorption spectra.

[18] F. Wang, M. Y. Sfeir, L. Huang, X. M. H. Huang, Y. Wu, J. Kim, J. Hone, S. O'Brien, L. E. Brus, and T. F. Heinz, Physical Review Letters **96** (16), 167401 (2006).

[19] P. Delaney, H. J. Choi, J. Ihm, S. G. Louie, and M. L. Cohen, Nature **391** (6666), 466 (1998); M. Ouyang, J. Huang, C. L. Cheung, and C. M. Lieber, Science **292** (5517), 702 (2001).

[20] M. Radosavljević, J. Lefebvre, and A. T. Johnson, Physical Review B **64** (24), 241307 (2001); J. Han and M. S. Strano, Materials Research Bulletin **58** (0), 1 (2014).

[21] S. Seppälä, E. Häkkinen, M. J. Alava, V. Ermolov, and E. T. Seppälä, Europhysics Letters **91** (4), 47002 (2010).

[22] T. Susi, A. Kaskela, Z. Zhu, P. Ayala, R. Arenal, Y. Tian, P. Laiho, J. Mali, A. G. Nasibulin, H. Jiang, G. Lanzani, O. Stephan, K. Laasonen, T. Pichler, A. Loiseau, and E. I. Kauppinen, Chemistry of Materials **23** (8), 2201 (2011).

[23] L. F. C. Pereira, C. G. Rocha, A. Latgé, J. N. Coleman, and M. S. Ferreira, Applied Physics Letters **95** (12), 123106 (2009).